\documentstyle[eqsecnum,aps,twocolumn]{revtex} 
\input psfig

\begin{document}
\draft
\preprint{ }
\title{Cosmological properties of a class of $\Lambda$ decaying cosmologies
}
\author{V. Silveira$^{1,3}$ and I. Waga$^{2,3}$}
\address{$^1$Universidade de Bras\'\i lia, International Centre of Condensed Matter Physics, C. P. 04513, 70919-970, Brazil. \\
$^2$Universidade Federal do Rio de Janeiro, Instituto de F\'\i sica, C. P. 68528, 21945-970, Brazil.\\
$^3$NASA/Fermilab Astrophysics Center, Fermi National Accelerator Laboratory,  Batavia, IL, 60510}
\date{\today}
\maketitle
\begin{abstract}
We investigate some properties of flat cosmological models with a 
$\Lambda$ term that decreases with time as $\Lambda \propto a^{-m}$ 
($a$ is the scale factor and $m$ is a parameter $0\leq m < 3$). 
The models are equivalent to standard cosmology with matter and 
radiation plus an exotic fluid with the equation of state 
$p_x = (m/3 -1)\rho_x$. We study the effect of the decaying $\Lambda$ term 
on the cosmic microwave background anisotropy and by using a 
seminumeric method we compute the angular power spectrum (up to $l=20$) for 
different values of $m$ and $\Omega_{m0}$. We also investigate 
the constraints imposed on the models by the magnitude-redshift test in which 
high-redshift type Ia supernovae (SNe Ia) are used as standard candles. 
We obtain the $95.4\%$, $90\%$, and $68\%$ confidence levels on the 
parameters $m$ and $\Omega_{m0}$ and compare them with those arising from
lensing statistics. Our analysis reveals that the SNe Ia constraints are stronger for low values of $m$ and $\Omega_{m0}$, while those from lensing statistics are more important for $m \stackrel{>}{\sim} 1$.
Models with $\Omega_{m0} \stackrel{>}{\sim} 0.2$ and 
$m \stackrel{>}{\sim} 1.6$ are in good agreement with the data.
\end{abstract}

\narrowtext

\section{Introduction}
\label{sec:level1}
Flat cosmological models with a cosmological constant are currently serious
candidates to describe the dynamics of the universe. There are three main
reasons for the present interest on these models. First, they can reconcile
inflation with dynamic estimates for the density parameter ($\Omega$).
Observations indicate $\Omega _{m0} = 0.2 - 0.4$ for matter that clumps on
scales $20 - 30h^{-1}$ Mpc, while inflationary models usually predict $\Omega
_{total}=1$. Second, these models, when normalized by data from the Cosmic Background Explorer (COBE) 
predict less power in the perturbation spectrum at 
small scales than standard cold dark matter (CDM), in accordance with observations \cite{efs}.
The third motivation for introducing a cosmological constant is the ``age
crisis.'' In flat models with $\Omega _{m0} = 1$ only if $h<0.57$ ($h$ is the
present value of the Hubble parameter in units of $100$ km/s Mpc$^{-1}$) 
is it possible to get theoretical ages for the universe that are higher than 
the lowest values ($\sim 12$ Gyear) estimated for the galactic globular cluster system \cite{age}. However, current estimates based on 
observations of Cepheids stars in the Virgo cluster and type Ia supernovae 
indicate higher values for $h$ \cite{tan}.  

In the past the cosmological constant was introduced and, with the improvement
of the observational data, later discarded. Now, however, the situation 
may change. By taking into account the vacuum
contribution to the energy-momentum tensor we can define an effective
cosmological constant ($\Lambda_{eff}$) that is the sum of two terms, 
the bare cosmological term and $8 \pi G \rho_{vac}$ ($\rho_{vac}$ 
is the vacuum energy density). From quantum field theory we should 
expect $\rho _{vac} \sim
M_{Pl}^4$ ($M_{Pl}$ is the Planck mass), or perhaps another energy  density
related to some spontaneous symmetry-breaking scale such as $M_{SUSY}$ or
$M_{weak}$ to the fourth power. The problem lies in that these values are
enormous when compared with astronomical bounds for $\rho _{\Lambda _{eff}}$. 
Extreme fine-tuning between $8 \pi G \rho _{vac}$ and the bare $\Lambda$ is 
necessary to make theory compatible with observations.  

One possible explanation for a small $\Lambda$ term is to assume that it is
dynamically evolving and not constant, that is, as the universe evolves 
from an earlier hotter and denser epoch, the effective
cosmological term also evolves and decreases to its present value \cite{lamb}. 
There are also strong observational motivations
for considering cosmological models with a decreasing $\Lambda$ term 
instead of a constant one. Usually in a dynamical-$\Lambda$ cosmological 
model, the
distance to an object with redshift $z$ is smaller than the distance to the
same object in a constant-$\Lambda$ model with the same value of the density
parameter. As we shall discuss in Sec. IV, this implies that constraints 
coming from lensing statistics and
from high redshift supernovae can be considerably weaker in these models \cite{blo}. 

Recently \cite{sil} we suggested a class of models in which $\Lambda$ decreases as
$\Lambda \propto a^{-m}$ [here $a$ is the scale factor of the Friedman-Robertson-Walker (FRW) metric and
$m$ is a constant ($0\leq m <3$)]. Although Chen and Wu \cite{che} gave some
interesting arguments favoring the special value $m=2$, it is
clear that the above functional dependence with the scale factor is only
phenomenological and does not come from particle physics first principles.
However, we believe it deserves further investigation for the following 
reasons.
First, these models generalize several other models present in the literature.
So, by investigating their properties we are studying at once the models 
they generalize. Second, since $m<3$ the universe age in these 
models is always larger than the age obtained in the standard Einstein--de 
Sitter 
cosmology and if $m<2$ the age is larger than the one we get in an open 
model with the same
$\Omega _{m0}$. This is an important aspect if we are interested in 
solving the ``age problem.'' 
Further, since $m <4$ the $\Lambda $ term is generally
not important during the radiation-dominated 
phase and nucleosynthesis proceed as in the 
standard model \cite{nuc}. Finally, the models are mathematically
simple and in most cases can be treated analytically.  

This paper is organized as follows. 
In Sec. II the basic equations of the models are obtained and our main assumptions presented. In Sec. III we
investigate the effect of the $\Lambda$ term on the cosmic microwave background (CMB) anisotropy and compute the angular power spectrum for small values of $l$ for different
values of the parameter $m$ and $\Omega_{m0}$. Constraints
on the models from high redshift SNe Ia and from lensing statistics are obtained in Sec. IV. 

\section{The models and the field equations}

In this paper we consider spatially flat, homogeneous, and 
isotropic cosmologies
with a time-dependent $\Lambda$ term: 
\begin{equation}
\Lambda = 8\pi G\rho _v=3\alpha a^{-m}. 
\end{equation}
\noindent 
Here the parameters $\alpha $ and $m$ are restricted to the range 
$0\leq m < 3$
, $\alpha \geq 0$ and the factor 3 was only introduced for 
mathematical convenience. We consider the cosmic fluid to be a
mixture of nonrelativistic matter and radiation 
($p_r=\frac {1}{3}\rho _{r}$)
with a perfect fluid energy momentum tensor,  
\begin{equation}
{{T}^\mu }_\nu =T_r^{}{}_{}^\mu {}_\nu +T_m^{}{}_{}^\mu {}_\nu =\hbox{ diag
}(\rho ,-p,-p,-p),  \end{equation}
where $\rho =\rho _r+\rho _m$ is the total energy density (radiation plus 
nonrelativistic matter) and $p=p_r$ is the thermodynamic
pressure. 

As in Ref. \cite{sil} we assume that vacuum decays only into relativistic
particles, such that the nonrelativistic matter energy momentum tensor is 
conserved ($\rho_m \propto a^{-3}$). The radiation energy density has two
parts: one conserved, $\Omega _{r0}H_0^2(a_0/a)^4$,  
$\Omega _{r0}=4.3 \times 10^{-5} h^{-2}$, and a second one, $\frac{3m\alpha
}{8\pi G(4-m)} a^{-m}$, which arises due to the vacuum decay. 
Here $a_0$ is the present value
of the scale factor and $H_0$ is the 
present value of
the Hubble parameter. In the following, subscripts $0$ will always indicate
present values. 

The Einstein equations for the models we are considering reduce to two
equations: namely,  
\begin{eqnarray}\left( \frac{\stackrel{.}{a}}a\right) ^2
&=&\Omega _{m0}H_0^2\left( \frac{a_0}a\right) ^3+
\Omega _{r0}H_0^2\left( \frac{a_0}a\right) ^4 \nonumber \\  &+&\Omega_{x0}H_0^2
\left( \frac{a_0}a\right) ^m \end{eqnarray}
\noindent and 
\begin{eqnarray}
\frac{\stackrel{..}{a}}a&=&-\frac 12\Omega _{m0}H_0^2
\left( \frac{a_0}a\right)^3-\Omega _{r0}H_0^2\left( \frac{a_0}a\right) ^4  \nonumber \\ &+&
\frac{(2-m)}2\Omega_{x0}H_0^2\left( \frac{a_0}a\right) ^m, \end{eqnarray}
\noindent
where $\Omega _{mo}$ is the matter density parameter and $\Omega _{x0}
=\frac{4\alpha H_0^{-2}a_0^{-m}}{(4-m)}$.

The above equations are quite general and apply for a broad spectrum 
of models. For instance, if $m=0$, the usual flat FRW model with a 
cosmological constant is obtained. If we take $m=2$, 
Eqs. (2.3) and (2.4) assume the same form of the Einstein equations for open 
models and also appear in some
string-dominated cosmologies \cite{vil}. Further, we would obtain the {\it same} 
Einstein equations if, instead of a cosmological term, we would have 
considered (beside conserved matter and radiation) an exotic x fluid with 
equation of state, $p_x=\left( \frac m3-1\right) \rho _x$ \cite{fry}. All we 
discuss here also applies for these cosmologies and to emphasize this point 
in the following we shall use the expression ``x component'' to 
interchangeably designate the $\Lambda$ term or the x fluid.
\narrowtext
\section{ The angular power spectrum}
The evolution of perturbations in the models we are considering may be 
studied considering two different phases \cite{sil}. For $a < a_{M}=a_0(\frac{\Omega
_{r0}}{\Omega _{x0}})^{\frac 1{4-m}}$, the energy  density of the 
universe is dominated by radiation and/or matter, and  the 
contribution from the x component may be neglected. During the second phase,
characterized by  $a>a_{M}$, the energy density of the universe is 
dominated by 
nonrelativistic matter and/or the x component. In this phase, 
the x component contribution becomes more and more important and 
the deviations from the usual matter-dominated cosmological model increase 
as the universe expands.

To describe the perturbation growth in these two phases, we consider the usual
linear perturbation theory. Each Fourier component of the perturbation,  
$\delta_{k}(t)$, grows independently of the other modes and may be related 
to the primordial power spectrum with the help of the transfer function:
\begin{equation}
\delta_{k}(t) = T(k) \delta_{k}( t_{i}),
\end{equation}
where $\delta_{k}(t_{i})$ is the primordial spectrum, usually taken to be 
\begin{equation}
\mid \delta _{k}(t_{i})\mid ^{2} = A k^{n}.
\end{equation}

In (3.1) $T(k)$ is the transfer function, which incorporates all deformations 
undergone by the $k$-mode perturbation and $t$ may be any time $t>t_{i}$. If 
the initial perturbation is adiabatic and $n=1$ we have a
primordial scale invariant adiabatic perturbation that is usually predicted 
by inflationary models. In fact, for flat models without a cosmological 
constant it can be shown that, in this case, the quantity $k^3 \mid \delta_{k}
 \mid ^2$ at 
$t=t_{enter}$, is independent of $k$ for any $k$. Here, $t_{enter}$ is the
instant when the perturbation crosses the Hubble radius. However, this is not
necessarily true if we have a cosmological term. In \cite{sil} we adopted a 
true scale invariant Harrison-Zeldovich spectrum for the models we are 
analyzing. Here we shall use the primordial scale invariant spectrum 
as defined above. 

The power spectrum is defined in the standard way by $P(k) = \mid 
\delta_{k}(t_{0})\mid ^{2}$ where $t_{0}$ is the present time. For 
flat models with $\Lambda =0$, and after radiation-matter equality
$(a>a_{eq})$, all modes, inside and outside the Hubble radius, evolve in the very
same way. So,  
$\delta_{k}(t)$, in fact, does not experience any deformation 
for  $t>t_{eq}$ and $P(k)$ reflects the power spectrum for any $t>t_{eq}$.
Since in our model  $a_{eq}<a_{M}$, the evolution during the first phase, 
dominated by radiation and matter, may be described by simply taking the 
standard power spectrum at $a=a_{M}$, with the transfer function 
as computed by Bond and Efstathiou \cite{bon}: 
\begin{equation}
T(k)=(1+(a k +(b k)^{3/2} +(c k)^{2})^{\nu})^{-\frac{1}{\nu}}
\end{equation} 
where $a= 6.4 (\Omega h^{2})^{-1}$ Mpc, 
$b=3.0 (\Omega h^{2})^{-1}$ Mpc, $c=1.7 (\Omega h^2)^{-1}$ Mpc, and $\nu=1.13$.

After radiation and x component equality, $a>a_{M}$, 
the increasingly importance of the x component
energy density must be taken into account. To describe the
perturbations after  
$a_{M}$, we consider the evolution of matter perturbations in the 
background dominated by matter and the x component. 
We shall use the
approximation that the x component is smooth. In this case it can be shown \cite{pad} 
that all modes, inside and outside the Hubble radius,
grow at the same rate as in a flat universe dominated by nonrelativistic
particles. So again no extra deformation is introduced in the power spectrum 
by the presence of the x component and, to study the multipole expansion of 
the power spectrum, we are allowed to use the transfer function given by Eq.
(3.3). 

Since one of our goals is to understand the effect of the decaying 
cosmological term (x component) on the  
CMB anisotropy, we
consider the relation between the mass density perturbation and the 
fractional
perturbation to the CMB temperature $\frac{\delta T}{T}$. Almost 30 years
ago, Sachs and Wolfe \cite{sac} obtained the expression relating
fluctuations in the gravitational potential on the last scattering surface 
with CMB temperature anisotropies on large angular scales 
($\theta \gg 1\deg$). For
a flat universe their formula can be written as \cite{pee}
\begin{eqnarray}
\frac{\delta T}{T}= & - &\left[\frac{a}{2}\frac{d D}{dt}\frac{\partial
k}{\partial x^\alpha}\gamma^\alpha \right]^{t_{ob}}_{t_{em}} -
\left[\frac{a}{2}\frac{d}{dt}\left(a\frac{dD}{dt}\right)k \right]
^{t_{ob}}_{t_{em}}  \nonumber \\ 
& + & \frac12 \int^{t_{ob}}_{t_{em}} dt k \frac{d}{dt}a
\frac{d}{dt}a\frac{dD}{dt}, 
\end{eqnarray}
where $D(t)$ describes the time dependence of the growing mode 
of $\delta_{m}$, $\delta_{m}( \vec{x}, t) = A(\vec{x}) D(t)$, 
$\gamma^\alpha$ is
the unit vector pointing along a null ray from the observer 
to the source, and
where $k(\vec{x})$ is given by 

\begin{equation} 
k(\vec{x}) = -\frac{1}{2 \pi} \int \frac{\delta_m(\vec{x'}, t)}
{D(t)}                      \frac{d^3 \,x'}{|\vec{x} - \vec{x'}|}    .
\end{equation}
The first term in the right-hand side (RHS) of (3.4) describes the  
anisotropy caused by the  relative motion source observer. 
The effect of the observer's motion may be systematically removed 
from the experimental data, and will not be considered any further.
The effect of the source motion is a Doppler contribution related 
to velocity perturbations on the last scattering surface and cannot be eliminated. The second term in Eq. (3.4) includes the 
usual Sachs-Wolfe effect (SW) and a constant part evaluated at 
the present epoch which does not contribute to the observed anisotropy.
Finally, there is the last term, usually called the integrated Sachs-Wolfe 
effect (ISW), which represents the redshift or blueshift of the photon energy 
caused by its traveling through regions of space with time-varying gravitational 
potential. Note that during the radiation-matter domination phase
the growing modes behave as \cite{sil} 
\begin{equation}
D(a)=D_{dec}(1+\frac{3a_{dec}}{2a_{eq}})^{-1}(1+ \frac{3a}{2a_{eq}}),
\end{equation}
where $a_{eq}=a_0\frac{\Omega_{r0}}{\Omega_{m0}}$ is the scale factor at
matter-radiation energy density equality. 
So, since in flat models, during the
radiation domination, we have $a\propto t^{1/2}$, and during matter 
domination
the scale factor grows as $a \propto t^{2/3}$, it is easy to see that the
contribution of the last term in Eq. (3.4) vanishes in both cases. This reflects the fact
that, during these eras, the gravitational potential is time independent.
However, during the x component domination, 
the gravitational potential will
not be constant and the ISW term will no longer be zero. In fact its 
contribution may be of the same order of the usual SW term for the lower 
modes in the harmonic expansion. In order to study the CMB anisotropies in 
the models described above,
we keep three contributions in the RHS of (3.4): SW, ISW, and Doppler.  

The full temperature correlation function is defined by
\begin{equation}
C(\alpha) = \langle \frac{\Delta T}{T}(\hat{n})
\frac{\Delta T}{T}(\hat{m})\rangle_{\hat{n} \cdot \hat{m}= \cos(\alpha)} 
\end{equation} 
where $\langle$ ... $\rangle$ denotes an average over all positions $\vec{x}$ and 
all directions $\hat{n}$, $\hat{m}$ separated by an angle $\alpha$.

To compute the average value in Eq. (3.7), we first analyze the ISW 
contribution \cite{jaf}. We have 
\begin{eqnarray}
&&C^{ISW}(\alpha)  =   \int_{a_{M}}^{a_{0}} da \int_{a_{M}}^{a_{0}} da' G(a)
G(a') \int \frac{d^3k}{(2 \pi)^{3}} \nonumber \\ &\times&\int \frac{d^3k^\prime}{(2
\pi)^{3}}\frac{\delta_k \delta^*_{k^\prime}}{k^2 {k^\prime}^2}  
\langle e^{i( \vec k^\prime - \vec k ). \vec x_{0}}\rangle_{x_0}\nonumber \\ &\times&  \langle e^{-i \vec k .
\hat{n}I_2(a,a_0) + i\vec k^\prime . \hat{m}I_2(a^\prime,a_0) } \rangle_
{\hat{n}. \hat{m}=\cos{(\alpha)}}.
\nonumber \\     
\end{eqnarray}
Here
\begin{equation}
I_2(a,a_0)= \frac{1}{H_{0}} \int_{a}^{a_{0}}\frac{dx}{\sqrt{\Omega_{m0} x +
(1-\Omega_{m0}) x^{4-m}}}, \end{equation}and 
\begin{equation}
G(a)=-\left( \frac{3D}{a}-D^\prime(3+\frac m2(\frac{a_d}{a})^{m-3})
\right)H_0^2 \Omega_{m0}a^2(\frac{a_0}{a})^3,
\end{equation}
where $a_d = a_0 ( \frac{\Omega_{m0}}{\Omega_{x0}})^{1/(3-m)}$ and $D' = dD/da$. 
Note that if $m=0$ and 
$D\propto a$ we obtain $G(a)=0$. Equation (3.8) was
obtained by making use of the field Eqs. (2.3) and (2.4), with
$\Omega_{r0}=0$, and the time evolution differential equation for $D$, 
\begin{equation}
\ddot{D} + 2 \frac{\dot{a}}{a} \dot{D} - 
4 \pi G \rho_{m} D = 0.
\end{equation}
By using that,
\begin{eqnarray}
\langle e^{i( \vec k^\prime - \vec k ). 
\vec x_{0}} \rangle_{x_0} & = &\frac{1}{V} 
\int d^3
x_0 e^{i( \vec k^\prime - \vec k ). \vec x_{0}} \nonumber \\  
&=&  \frac{(2\pi)^3}{V}\delta^3 (\vec k^\prime - \vec k )
\end{eqnarray}
and
\begin{eqnarray}
&&\langle e^{-i \vec k . \hat{n}I_2(a,a_0) + 
i\vec k^\prime . \hat{m}I_2(a^\prime,a_0)
} \rangle_{\hat{n}. \hat{m}=\cos{(\alpha)}} 
\nonumber \\& = & j_0\left[ k\left({I_2(a,a_0)}^2 +I_2(a^\prime,a_0)^2
\right. \right. \nonumber \\  & &- \left. \left.
2I_2(a,a_0)I_2(a^\prime,a_0
)\cos(\alpha)\right)^{1/2}\right],  
\end{eqnarray}
where $j_0(x)=\frac{\sin(x)}{x}$, 
we reduce (3.8) to 
\begin{eqnarray}
&&C^{ISW}(\alpha)=  \int_0^\infty \frac{dk}{2 
\pi^{2}}\frac{P(k)}{k^2}\int_{a_{M}}^{a_{0}} da \int_{a_{M}}^{a_{0}} da' 
G(a) G(a')  \nonumber \\
&\times& j_0\left[ k\left({I_2(a,a_0)}^2 +I_2(a^
\prime,a_0)^2 \right. \right.\nonumber \\
&-& 2 \left. \left. I_2(a,a_0)I_2(a^\prime,a_0
)\cos(\alpha)\right)^{1/2}\right], 
\end{eqnarray}
where $P(k)= A k T(k)^2$. 
Analogously we obtain
\begin{eqnarray}
&&C^{SW}(\alpha) = \left(\frac{3H_{0}^{2} 
\Omega_{m0} D_{dec}a_0^3}{4a_{eq}(1+ \frac32
a_{dec}/a_{eq})} \right)^2 \times 
\nonumber \\ & &\int_{0}^\infty \frac{dk}
{2 \pi^{2}}\frac{P(k)}{k^2} j_{0}[2 k \left(I(a_{dec},a_0)\right) 
\sin(\alpha/2)], \end{eqnarray}where $I(a_{dec},a_0)=I_1(a_{dec},a_M)+
I_2(a_M,a_0)$, with
\begin{equation}
I_1(a_{dec},a)= \frac{1}{H_{0}} \int_{a_{dec}}^{a}\frac{dx}
{\sqrt{\Omega_{m0}x
+ (1-\Omega_{m0}) }}. 
\end{equation}
The Doppler term is given by
\begin{eqnarray}
&&C^{Dop}(\alpha) = {a_{dec}}^{2} \dot{D}_{dec}^2 \int_0^\infty  
\frac {dk}{2 \pi^{2}} \frac{P(k)}{k^2} \times \nonumber \\&& 
\frac{d}{dx} \frac{d}{dy} \left(j_{0}[k \sqrt{x^{2} + y^{2} - 
2x y\cos(\alpha)}]\right)_{x=y=I(a_{dec},a_0)}.\nonumber \\
\end{eqnarray}
The cross terms reduce to
\begin{eqnarray}
&&C^{ISW-SW}(\alpha) 
= \nonumber \\&-&  \frac{3H_{0}^{2} \Omega_{m0} D_{dec}a_0^3}{2a_{eq}(1+ \frac32 
a_{dec}/a_{eq}}) \int_{a_{M}}^{a_{0}} da G(a)\int_0^\infty  
\frac{dk}{2 \pi^{2}} \frac{P(k)}{k^2} \nonumber \\&\times & j_{0}[ k 
\left({I(a_{dec},a_0)}^2 + {I_2(a,a_0)}^2 \right. \nonumber \\ &-& \left.        2 I(a_{dec},a_0) 
I_2(a,a_0)\cos(\alpha)\right)^{1/2}]
\end{eqnarray}
\begin{eqnarray}
&&C^{Dop-SW}(\alpha)=(\frac{3H_{0}^{2} \Omega_{m0} \dot{D}_{dec}
D_{dec}a_0^3}{(\frac{a_{eq}}{a_{dec}}+ \frac32 )}) 
 \nonumber \\ 
& \times &\int_0^\infty  \frac{dk}{2 \pi^{2}} \frac{P(k)}{k^2}\frac{d}{dx}\left(j_{0}[k \left(x^{2} + I(a_{dec},a_0)^{2} \right. \right. \nonumber \\ &-& \left. \left.2 x
I(a_{dec},a_0) \cos(\alpha)\right)^{1/2}]\right)_{x=I(a_{dec},a_0)} 
\end{eqnarray}
\begin{eqnarray}
&&C^{Dop-ISW}(\alpha)= -2 a_{dec} \dot{D}_{dec}\int_{a_{M}}^{a_{0}} da 
G(a)  
 \nonumber \\ &\times&\int_0^\infty  \frac{dk}{2\pi^{2}} \frac{P(k)}{k^2}\frac{d}{dx}\left(j_{0}[k \left(x^{2}+I(a,a_0)^{2}\right. \right. \nonumber \\ &-& \left. \left. 
2 x I(a,a_0)
\cos(\alpha)\right)^{1/2}]\right)_{x=I(a_{dec},a_0)}
\end{eqnarray}
The total contribution to (3.7) may be written as
\begin{eqnarray}
C(\alpha)&=& C^{ISW}+C^{SW}+C^{Dop}+C^{ISW-SW}\nonumber \\
&+&C^{Dop-SW}+C^{Dop-ISW}
\end{eqnarray}

It is convenient to expand the temperature correlation function 
in angular multipoles, using Legendre Polynomials $P_{l}(\cos(\alpha))$:
\begin{equation}
C(\alpha) = \sum_{l=0}^{\infty} \frac{2l+1}{4 \pi}            
P_{l}(\cos(\alpha)) C_l\end{equation}
and 
$C_{l}= 2 \pi \int_{-1}^{1} d \cos(\theta) P_{l}(\cos(\theta)) C(\theta)$.
With the help of the relation \,\, $j_{0}[k \sqrt{r^{2} + 
q^2 - 2 r q \cos(\theta)}] =  
\sum_{n=0}^{\infty} (2n+1) P_{n}(\cos(\theta)) j_{n}[kr] j_{n}[kq]$, where $j_l[x]$ is a spherical Bessel function, the
integrals in  
the contributions for $ C_l$ decouple and the angular power spectrum becomes
\begin{eqnarray}
C_{l} &=& C_{l}^{ISW} + C_{l}^{SW} + C_{l}^{Dop}+C_{l}^{ISW-SW}\nonumber \\
&+&C_{l}^{Dop-SW}+ C_{l}^{Dop-ISW}
\end{eqnarray}
with
\begin{equation}
C_{l}^{ISW}=\frac{2}{\pi} \int_0^\infty \frac{dk}{k^2} P(k)( X_{l}(k))^2 
\end{equation}    where
\begin{equation}
X_{l}(k) =  \int_{a_{M}}^{a_{0}} da G(a) j_l[ k I(a,a_0)],        
\end{equation}
\begin{eqnarray}
C_{l}^{SW} &=&\frac2\pi \left(\frac{3H_{0}^{2} \Omega_{m0}
D_{dec}a_0^3}{4a_{eq}(1+ \frac32 a_{dec}/a_{eq})} \right)^2 \nonumber \\
&\times&\int_0^\infty 
\frac{dk}{k^{2}} P(k) j_{l}^{2}[kI(a_{dec},a_0)],
\end{eqnarray}
\begin{eqnarray}
C_{l}^{Dop}&=& \frac2\pi{a_{dec}}^{2} \dot{D}_{dec}^2\nonumber \\
&\times&
\int_0^\infty \frac{dk}{k^2} P(k) \left(\frac{d}{dx}\left(j_{l}[k
x]\right)_{x=I(a_{dec},a_0)}\right)^{2}, 
\end{eqnarray}
\begin{eqnarray}
C_{l}^{ISW-SW}&=&- \frac{3H_{0}^{2} \Omega_{m0} D_{dec}a_0^3}{\pi a_{eq}(1+
\frac32 a_{dec}/a_{eq})} \nonumber \\
&\times&\int_0^\infty \frac{dk}{k^2} 
P(k) j_{l}[kI(a_{dec},a_0)] X_{l}(k), 
\end{eqnarray}
\begin{eqnarray}
&&C_{l}^{Dop-SW}= (\frac{3H_{0}^{2} \Omega_{m0} \dot{D}_{dec}
D_{dec}a_0^3}{\pi(\frac{a_{eq}}{a_{dec}}+ 
\frac32 )}) \int_0^\infty \frac{dk}{k^2} P(k) \nonumber \\
&\times&j_{l}[k I(a_{dec},a_0)]\frac{d}{dx}\left(j_{l}[k x]
\right)_{x=I(a_{dec},a_0)},
\end{eqnarray}
and
\begin{eqnarray}
&&C_{l}^{Dop-ISW}= - \frac{4 a_{dec} \dot{D}_{dec}}{\pi}\nonumber \\
&\times&                 
\int_0^\infty  \frac{dk}{k^2} P(k)  X_{l}(k)\frac{d}{dx}\left(j_{l}[k
x]\right)_{x=I(a_{dec},a_0)}.  
\end{eqnarray}

Using the expressions above, we compute the angular power spectrum 
for different values of $m$ and $\Omega_{m0}$. 
Calculations are done in a semianalytical way through the following steps.
To make the integration in $X_{l}(k)$ [Eq. (3.25)], we change variables. By 
defining $w=I(a, a_0)$, and by using an interpolated polynomial, 
we rewrite the integral in the new  $w$ variable as
\begin{equation}
X_{l}=-\int_{0}^{w_{M}} dw\; CI(w)\;  j_l[k w].
\end{equation}    
Here $CI(w)$ is an interpolated polynomial, which replaces the function 
$C(w)= \frac{d}{dw}(I^{(-1)}(w)) G(I^{(-1)}(w))$, 
where $I^{(-1)}(w)$ stands
for the  inverse of $w=I(a, a_0)$. 
The precision of this interpolation may be estimated by
comparing $CI(w)$ and the original function $C(w)$. In the models we consider,
this comparison shows that using the seventh-order interpolated polynomial,
$CI(w)=\sum_{n=0}^{7} a_{n} w^{n}$,   
introduces an error always smaller than $0.07\%$.
With the help of this approximation, 
the resulting seven integrals in $X_{l}$ can be analytically computed. 
The form of these expressions remains the same for all models that we want to consider. Changing among models only changes the
numerical values of $a_{n}$ together with the upper limit in the 
integral, $w_{M}$. Once we have calculated these auxiliary functions $X_l(k)$, we may compute the coefficients in Eqs. (3.24), (3.28), and (3.30) by direct integration in $k$. In the limit 
$k \rightarrow 0$, the integrand is well behaved. For large values of 
$k$, the integrand decreases very fast. However, while decreasing, 
these integrands oscillate wildly, making the integral procedure 
slower and slower.
To avoid this technical difficulty, we truncate the integral at some value of
$k$, and we certify ourselves that the truncation is affecting our results 
in a controlled way. 
For values of $k$ larger than the truncation value, we compute the integrals 
replacing the oscillating $j_{l}(k I)$ by an envelope function, and the 
result of these integrals is taken as an upper bound of the error 
introduced by the truncation in $k$. This error grows
steadily with $l$, starting with $0.03\%$ for $l=2$ and growing up to 
the range $2-3\%$ for $l=20$ and different values of $m$.  

The numerical results obtained with the method described above are plotted in  Figs. $1$ and $2$.

\begin{figure} \hspace*{0.1in}
\psfig{file=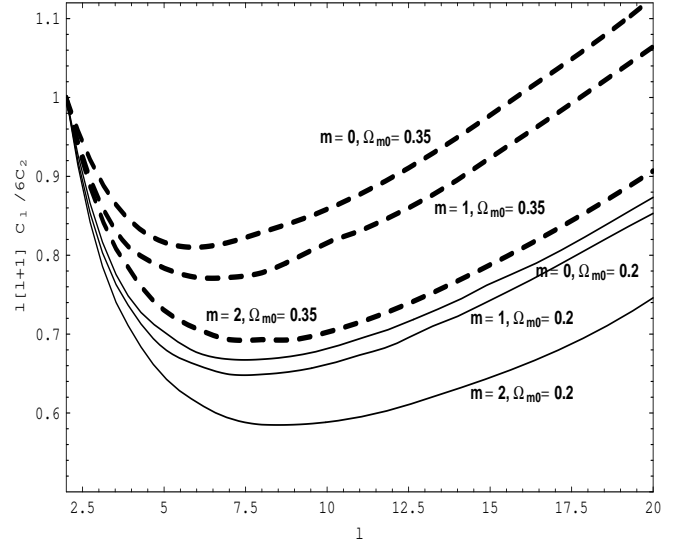,height=6.5cm,width=7.8cm}
\vspace*{0.4in}
\caption{The angular power spectrum $l(l+1)C_{l}/6C_2$ $\times$ $l$ is shown for different values of $m$ and $\Omega_{m0}$.}
\end{figure}

\begin{figure} \hspace*{0.1in}
\psfig{file=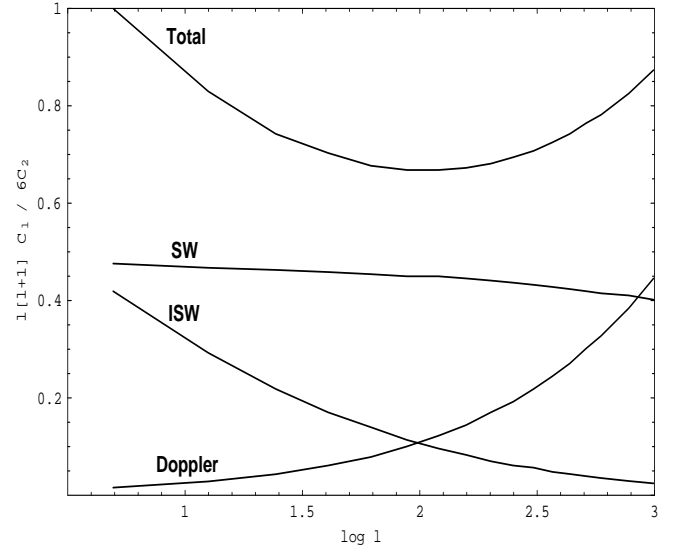,height=6.5cm,width= 7.8cm}
\vspace*{0.4in}
\caption{The Doppler, the Sachs-Wolfe, and the integrated Sachs-Wolfe contributions to the angular power spectrum are shown. For the figure we considered the special model for which $m=2$ and $\Omega_{m0}=0.2$.}
\end{figure}

In Fig. $2$ we plot for $m=2$ and $\Omega_{m0}=0.2$ the separate 
contributions from the 
Sachs-Wolfe, Doppler and integrated Sachs-Wolfe effects. 
Crossed terms are not shown.
Basically, in the range $2 \leq l \leq 20$, the Sachs-Wolfe contribution 
remains always important, with a slight decrease. 
The integrated Sachs-Wolfe is large
for small values of $l$, but soon it becomes unimportant, reducing to just
$2\%$ of $C_{l}$ at $l=20$. On the other hand, the Doppler 
contribution starts very small and grows fast to become the 
dominant contribution at $l=20$.  

For the
analyzed models, the curves $l(l+1)C_{l}/6C_2$ $\times$ $l$ in Fig. $1$,
show a common feature, a minimum
for small values of $l$. We see that, for fixed $\Omega_{m0}$ and $h$,  the 
position of this minimum changes with
$m$, becoming deeper as $m$ grows larger. This feature reflects the fact that the contribution from the ISW effect is larger for larger values of the parameter $m$. 
We also observe that for fixed $m$ and $h$ and different values of the matter content, the minimum grows deeper
for smaller values of $\Omega_{m0}$. Again this feature agrees with the fact that it is the x component term which brings up the nonzero ISW effect,
which, by its turn, makes the minimum deeper. So, since $\Omega_{total}=1$, 
less nonrelativistic matter means
more x component contribution, more ISW effect, and consequently deeper minimum.

\narrowtext
\section{Constraints from high-redshift type Ia Supernovae and lensing statistics}
The Supernovae Cosmology Project is an ongoing program to systematically 
search and study high-z supernovae. 
In a recent report \cite{per} Perlmutter {\it et al.} analyzed seven 
SNe (with redshift $z=0.35 - 0.46$), of more than 28 supernovae discovered, 
and obtained constraints on cosmological parameters, specially on the 
cosmological constant. Their preliminary result, 
$\Omega_{\Lambda} < 0.51$ at the $95\%$ confidence level, 
strongly constraints models for the universe with a cosmological constant. 
In this section we use their observational results and adapt their 
procedure to constraint the class of models described in Sec. II.

The basic idea is to use type Ia supernovae as standard candles 
for the classic magnitude-redshift test. As in \cite{per} we express the 
apparent bolometric magnitude $m(z)$ as
\begin{equation}
m(z) = {\cal{M}} + 5 \;\log d_l(z, \Omega_{m0}, m)
\end{equation}
where in our case the luminosity distance (in units of $H_0^{-1}$) 
is given by,
\begin{eqnarray}
&&d_l(z, \Omega_{m0}, m)= c (1+z)\nonumber \\
&\times& \int_0 ^z\frac{dz}{\sqrt{(1+z)^3 \Omega_{m0} 
+ (1+z)^m (1-\Omega_{m0})}}
\end{eqnarray}
In Eq. (4.1) 
\begin{equation}
{\cal{M}} = \mbox{M} - 5 \;\log H_0 + 25
\end{equation}
is the ``zero point'' magnitude (or Hubble intercept magnitude), 
which is estimated from the apparent magnitude and redshift of low-redshift 
($z < 0.1$) SNe Ia. The nearby supernovae data set used by Perlmutter {\it et al.} in the determination 
of $\cal{M}$ were those 18 SNe Ia, discovered by the Calan/Tololo Supernovae 
Search \cite{ham} for which the first observations were made no later than 5 days after 
maximum.

Although SNe Ia are very similar explosion events, 
it is now known that they do not constitute a completely homogeneous class. 
Recently progress was made in the study of their inhomogeneities. 
Phillips \cite{phi} showed that there is a correlation between absolute 
magnitudes (M) at maximum light and the initial decline parameter 
$\Delta m_{15}$, the B-magnitude decline in the first 15 days after 
maximum. By studying the Calan/Tololo type Ia supernovae, Hamuy {\it et al.} 
\cite{ham} confirmed the existence of the correlation suggested by Phillips and obtained a prescription 
for correcting observed B-magnitudes to make them comparable to an arbitrary 
``standard'' SNe Ia light curve of width $\Delta m_{15} =1.1$. After adding 
the correction to the SNe Ia low-z set they reduced the magnitude dispersion 
from $0.26$ to $\sigma_{\mbox{M}_B,corr}^{Hamuy}=0.17$. Perlmutter 
{\it et al.} did the same for the seven high-z supernovae and reduced 
the dispersion from $0.27$ to $0.19$ mag. 
\begin{figure} \hspace*{0.1in}
\psfig{file=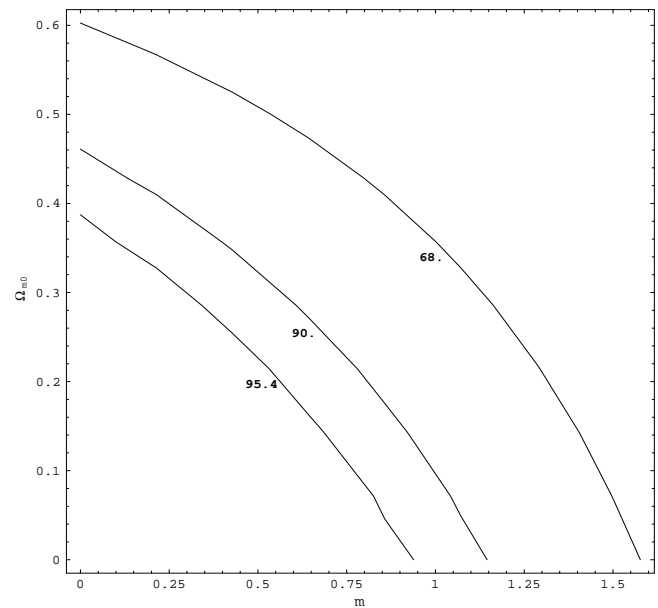,height=7.5cm,width=7.8cm}
\vspace*{0.4in}
\caption{Constraints imposed on the models by the magnitude-redshift test in which high-z SNe Ia are used as standard candles. The $95.4\%$, $90\%$, and $68\%$ confidence levels on the 
parameters $m$ and $\Omega_{m0}$ are shown in the figure.}
\end{figure}

In our computations we follow \cite{per} and use the corrected B-magnitude 
intercept at $\Delta m_{15}=1.1$ mag, ${\cal{M}}_{B,corr}^{\{1.1\}}= 
-3.32 \pm 0.05$, and consider only those 5 supernovae that have 
$\Delta m_{15}$ values in the range $0.8 - 1.5$ mag, which is the 
range of values investigated in the 18 low-z supernovae dataset. 
To construct the $\chi^2$ we used the data points 
outer error bars of Perlmutter {\it et al.}, which are obtained by adding in quadrature the inner error 
bars of $m_{B,corr}$ (the apparent B-magnitude after width-luminosity 
correction) to  $\sigma_{\mbox{M}_B,corr}^{Hamuy}$.

In Fig. $3$ we plot the $95.4\%$, $90\%$, and $68\%$ confidence levels for 
the parameters $m$ and $\Omega_{m0}$. We see that for the interesting range 
$\Omega_{m0}\stackrel{>}{\sim}0.2$, models with $m \stackrel{>}{\sim} 1.3$ 
are in good agreement with the data. The goodness-of-fit (as defined 
in \cite{per}) for the considered case is $0.59$. 

If we fix $m=0$ we recover the result of Perlmutter {\it et al.} for 
constant $\Lambda$, $\Omega_{\Lambda}= 0.06^{+0.28}_{-0.34}$, with 
$\Omega_{\Lambda}<0.51$ at the $95\%$ confidence level (one tail). Another interesting 
case is $m=2$. In this case we again have to consider one degree of freedom 
in $\Delta \chi^2 = \chi^2 - \chi^2_{min}$, and we find  $\Omega_{m0}= 0.83^
{+0.82}_{-0.69}$ ($1 \sigma$) with goodness-of-fit equal to $0.75$. So, models with $m=
2$ and $\Omega_{m0} > 0.14$ are also in good agreement with the data.

It is interesting to compare the above constraints with those arising from lensing statistics. In Fig. $4$ we plot contours of constant likelihood
($95.4\%$, $90\%$, and $68\%$) for the models, arising from lensing statistics.
We used data from the HST Snapshot survey, the Crampton survey, the Yee survey, the ESO/Liege survey, The HST GO observations, the CFA survey, and
the NOT survey \cite{mao}. We considered 859 ($z > 1$) quasars plus 5 lenses and to obtain the contours we adapted from \cite{blo} the procedure denoted there as ``A2''.
\begin{figure} \hspace*{0.1in}
\psfig{file=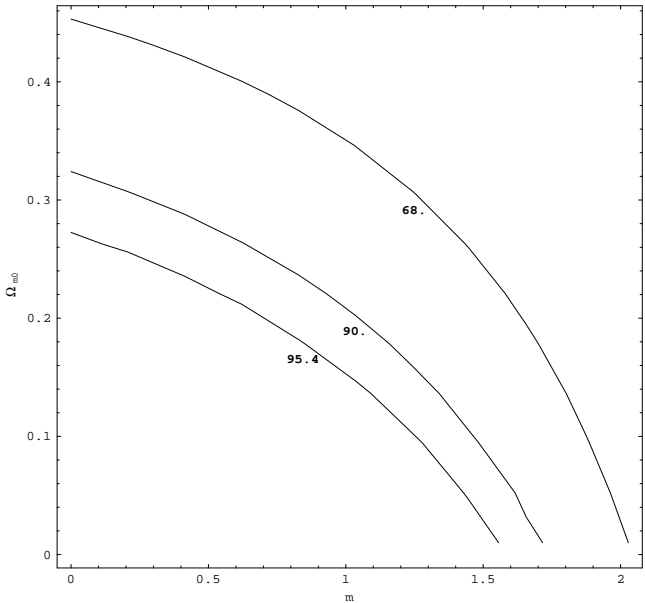,height=7.5cm,width=7.8cm}
\vspace*{0.4in}
\caption{Contours of constant likelihood ($95.4\%$, $90\%$ and $68\%$) arising from lensing statistics are shown.}
\end{figure}

By comparing Fig. $3$ with Fig. $4$ we see that the SNe Ia constraints are 
stronger for low values of $m$ and $\Omega_{m0}$ ($m <1 $ and $\Omega_{m0} <
0.4$) while the lensing ones are more important for larger values of the parameter $m$ ($m > 1$). From Fig. $4$ we observe that models with $\Omega_{m0} \stackrel{>}{\sim} 0.2$ and 
$m \stackrel{>}{\sim} 1.6$ are in good agreement with the data.

\acknowledgments
It is a pleasure to thank Scott Dodelson, Josh Frieman, and Albert Stebbins for several useful discussions. Special thanks are due to Saul Perlmutter for kindly answering our questions and Chris Kochanek for sending us data on lensing surveys.
This work was supported in part by the Brazilian agency 
CNPq and by the DOE and NASA at Fermilab through the grant NAGW-2381.


\begin{references}
\bibitem{efs}G.Efstathiou, W. J. Sutherland, and S. J. Maddox , Nature (london){\bf 348}, 705 (1990).
\bibitem{age} D. A. VandenBerg, M. Bolte, and P. B. Stetson, Annu. Rev. Astron. Astrophys. {\bf 34}, 461 (1996); see also, B. Chaboyer {\it et al.}, astro-ph 9706128, with new results based on recent Hipparcos parallax measurements which indicate younger ages ($\sim 10$ $\%$ smaller) for the oldest globular clusters. If these results are confirmed the age problem will possibly be weakened. However, it is still early to say that the age crisis has gone away (see, e.g., W. L. Freedman, in astro-ph 9706072 for a recent discussion on this and related topics). 
\bibitem{tan}N. R. Tanvir {\it et al.}, Nature (London) {\bf 377}, 27 (1995); W. Freedman, 
astro-ph/9612204; A. G. Riess, W. H. Press, and R. Kirshner, Ap. J. {\bf 438},
L17 (1995); M. Hamuy {\it et al.}, Astron. J. {\bf 112}, 2398 (1996).   
\bibitem{lamb}See, e.g., M. Ozer and M. O. Taha, Nucl. Phys. {\bf B287}, 776 (1987); K. Freese {\it et al.}, Nucl. Phys. {\bf B 287}, 797 (1987); M. Reuter and C. Wetterich, Phys. Lett. {\bf B 188}, 38 (1987); B. Ratra and P. J. E. Peebles, Phys. Rev. {\bf D 37}, 3407 (1988); 
I. Waga, Ap. J. {\bf 414}, 436 (1993); J. A. Frieman {\it et al.}, Phys. Rev. Lett. {\bf 75}, 2077 (1995); K. Coble, S. Dodelson, and J. A. Frieman, Phys. Rev. {\bf D 55}, 1851 (1997).
\bibitem{sil}V. Silveira and I. Waga, Phys. Rev. {\bf D 50}, 4890 (1994).
\bibitem{che}W. Chen and Y. S. Wu, Phys. Rev. {\bf D 41}, 695 (1990). 
\bibitem{nuc}For the models considered in this paper $\Omega_{\Lambda}$ is negligibly small during nucleosynthesis. However, this is not necessarily true for all $\Lambda$-decaying models present in the literature. For instance, for a class of models in which $\Omega_{\Lambda} = \beta $ = constant,  Freese {\it et al.} \cite{lamb} showed that standard primordial nucleosynthesis constraints the parameter $\beta$ to be $\beta \stackrel{<}{\sim} 0.1$.
\bibitem{vil}A. Vilenkin, Phys. Rev. Lett. {\bf 53}, 1016 (1984);
D. N. Spergel and U. L. Pen, astro-ph/9611198; L. M. A. Bettencourt,
P. Laguna and, R. A. Matzner, astro-ph/9612350.
\bibitem{fry}J. N. Fry, Phys. Lett. {\bf B 158}, 211 (1985); 
V. Sahni, H. A. Feldman and A. Stebbins, Ap. J. {\bf 385}, 1 (1992); 
H. A. Feldman and A. E. Evrard, Int. J. Mod. Phys. {\bf D 2}, 113 (1993); 
J. Stelmach and M. P. Dabrowski, Nucl. Phys. {\bf B 406}, 471 (1993);
H. Martel, Ap. J. {\bf 445}, 537 (1995); P. J. Steinhardt, Nature (London), {\bf 382}, 768 (1996); M. S. Turner and M. White, astro-ph/9701138.
\bibitem{bon}J. R. Bond  and G. Efstathiou, Ap. J. {\bf 285}, L45 (1987);
G. Efstathiou, Proc. Natl. Acad. Sci. USA {\bf 90}, 4859 (1993).
\bibitem{pad}The evolution of modes bigger than the Hubble radius is described by the general relativistic equations [see, e. g., T. Padmanabhan, {\it Structure Formation in the Universe}, (Cambridge: Cambridge University Press), p. 141 - 146 (1993)]. Since nonrelativistic matter is conserved, and assuming that the x component is smooth the relativistic equation for $\delta \dot{H}$ may be written as an equation for $\delta_{m}(\vec{x}, t)$, giving us the same Newtonian equation used to describe the evolution of modes inside the horizon.
\bibitem{sac}R. G. Sachs and A. M. Wolfe, Ap. J. {\bf 147}, 73 (1967).
\bibitem{pee}P. J. E. Peebbles, {\it Principles of Cosmology} (Princeton: Princeton University Press, 1993).
\bibitem{jaf}A. H. Jaffe, A. Stebbins, and J. A. Frieman, Ap. J. 
{\bf 420}, 9 (1994).
\bibitem{per}S. Perlmutter {\it et al.}, Ap. J. {\bf 483}, 565, (1997). 
\bibitem{ham}M. Hamuy {\it et al.}, A. J. {\bf 109}, 1 (1995); 
M. Hamuy {\it et al.}, A. J. {\bf 112}, 2391 (1996);
\bibitem{phi}M. M. Phillips, Ap. J. {\bf 413}, L105 (1993). 
\bibitem{blo}L. F. Bloomfield Torres and I. Waga, Mon. Not. R. Astron. Soc. 
{\bf 279}, 712 (1996).
\bibitem{mao}D. Maoz {\it et al.}, Ap. J. {\bf 409}, 28 (1993); D. Crampton ,R. D. McClure, and J. M. Fletcher, Ap. J. {\bf 392}, 23 (1992); H. K. C. Yee , A. V. Filipenko, and D. H. Tang, A. J. {\bf 105}, 7 (1993); A. J. Surdej {\it et al.} {\bf 105}, 2064 (1993); E. E. Falco, in {\it Gravitational Lenses in the Universe}, edited by J. Surdej, D. Fraipont-Caro, E. Gosset, S. Refsdal, and M. Remy (Liege: Univ. Liege), 127 (1994); C. S. Kochanek, E. E. Falco, and R. Shild, Ap. J. {\bf 452}, 109 (1995); A. O. Jaunsen {\it et al.}, A \& A {\bf 300}, 323 (1995). 

\end{references}
\end{document}